# Information Networks Secured by the Laws of Physics

L.B. Kish [†], F. Peper [††]

**SUMMARY** In this paper, we survey the state of the art of the secure key exchange method that is secured by the laws of classical statistical physics, and involves the Kirchhoff's law and the generalized Johnson noise equation, too. We discuss the major characteristics and advantages of these schemes especially in comparison with quantum encryption, and analyze some of the technical challenges of its implementation, too. Finally, we outline some ideas about how to use already existing and currently used wire lines, such as power lines, phone lines, internet lines to implement unconditionally secure information networks.

## 1. Introduction: quanta of security and insecurity

This paper shall address some key arguments and features related to the recent secure communication scheme utilizing the laws of classical physics with the specific goal to make secure wire-based information networks possible.

Today, in the internet era, data communication security is becoming one of the most important aspects of everyday life; it is the essential requirement to function while our computer is connected to the internet. Any one of the following cases means absolute security:

a) The eavesdropper (Eve), basically, cannot physically access the information channel. Tamper-resistant line methods belong to this case (for example, see [1]).

b) Eve has access and can do measurements on the channel but the communicating parties (Alice and Bob) already have a shared secret key, which they can use for encryption.

c) Eve has access and can do measurements on the channel but the laws of physics do not allow extracting the communicated information from the measurement data.

d) Eve can execute an invasive attack on the line and she can potentially extract the communicated information however, when that happens, she disturbs the channel so much that Alice and Bob discover the attack.

Tamper resistant lines [1] have been demonstrated by the idea that the energy dissipation processes are occurring at the nanoscale, not at the macroscale which is easily tampered with. However, such a secure signal transfer is demonstrated in nanoscale or nanophotonic devices [2], not in Internet-scale system yet. Thus usually situation b) is relevant for today's secure communications. Consequently, the "*key question*" of security is: How to share securely a secret key between Alice and Bob.

Accordingly, in today's software-based secure communications, before the secure data exchange can start, the two communicating parties (Alice and Bob) must generate and share a joint secret (secure) encryption key through the communication channel while the eavesdropper (Eve) is supposedly monitoring the related data (Figure 1) [3]. This is a mathematically impossible task with current software methods. If a sufficient calculation power is available for Eve, she can extract the secure key and decrypt the communicated data with a reasonable speed thus current software-based key distribution scheme are only "computationally safe". New algorithms and computing solutions are continuously researched, including quantum computing [4] and noise-based logic [5], therefore today's software-based secure communication contains a potential time bomb.

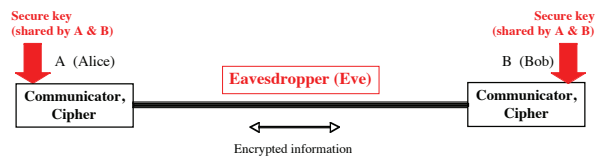

**Figure 1.** Alice and Bob must generate and share a joint secure key through the communication channel while the eavesdropper (Eve) is monitoring the related data [3]. This is an impossible task with currently used software-based methods.

Quantum key distribution [6] (Stephen Wiesner 1970's; Charles H. Bennett and Gilles Brassard 1984; Artur Ekert 1990) has offered a solution, which is claimed to

---

be unconditionally secure, an expression indicating that no reservations about Eve's computational limits are made. The information bits are carried by single photons (Figure 2). Here the *no-cloning-theorem* of quantum physics is the theoretical foundation of security. It means that a single photon cannot be copied without noise (error). If Eve captures and measures the photon, it gets destroyed and she must regenerate and reinject it into the channel otherwise this bit will be considered invalid by Alice and Bob. However, due to the no-cloning rule, while Eve is doing that, she introduces noise and the error rate in the channel will become greater than without eavesdropping. Therefore, by evaluating the error statistics, Alice and Bob will discover the eavesdropping. And here is the weak point: *statistics*. First of all, error statistics based eavesdropper detection means that a large number of bits must be exchanged to see deviations from the regular statistics. The security of a single bit or a few bits is very poor. Secondly, statistics based decision means that it can be a wrong decision and the probability of correct decision, that is the probability that the key exchange is indeed secure, is determined on the invested resources and it can never be 100%. For example, for the quantum *intercept and resend attack* [6], the probability $P$ of uncovering Eve, $P = 1 - 0.75^N < 1$, where $N$ is the number of bits in the key. This is one example, where the classical physical noise-based secure key exchange discussed below is superior to quantum encryption.

Moreover, quantum communicators are vulnerable to the advanced type of the man-in-the-middle-attack, where Eve breaks the channel and installs two communicators. With one of them she will communicate with Alice while pretending that she is Bob, and with the other one she will communicate with Bob while pretending that she is Alice. This is another important example, where the noise-based secure wire communicator discussed below is superior to quantum encryption.

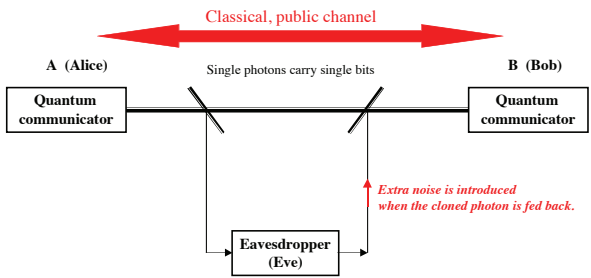

**Figure 2.** Generic quantum communication arrangement [3]. To detect the eavesdropper, a statistics of bit errors must be built. That requires a sufficiently large number of bits.

Many quantum key exchangers have reportedly been built [6], up to the range of 200 km. Most of them are working through optical fibers and some of the most advanced and secure ones are able to communicate via air [6]. However, the security claim of quantum encryption is mostly theoretical and experimental efforts to crack quantum encryption have attracted only limited efforts to date [7-9]. Nevertheless, these quantum attacks, which are based on the non-idealities of building elements and called "*quantum hacking*" have been reported to be extremely successful [6-15]. This indicates that much research and tests are yet to be carried out with regard to the security of communications, and that sufficient security and cost-effectiveness of quantum-based methods have not yet been satisfactorily established [7-15].

## 2. Classical physics offers a robust security solution

The main topic of this paper is the a secure key exchange scheme introduced in 2005 [16,17] and built/demonstrated in 2007 [18], which is utilizing the robustness of classical information, stochasticity, and the laws of classical physics to provide a security that, in several aspects, looks superior to that of quantum encryption. This scheme was named by its creators as Kirchhoff-loop-Johnson(-like)-Noise (KLJN) scheme while on the internet has widely been nicknamed as Kish cypher, Kish cipher, Kish's scheme, etc.

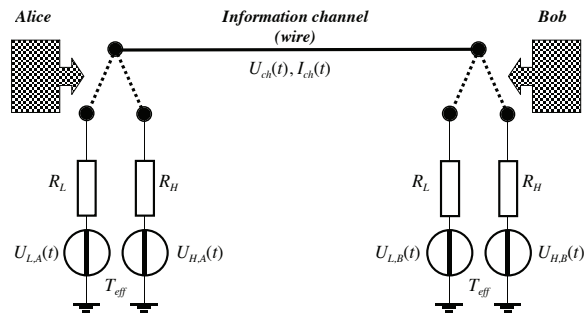

**Figure 3.** The core of the KJLN secure key exchange system. In the text below, the mathematical treatment utilizes the power density spectra of the voltages and currents shown in the figure.

The KLJN scheme is a statistical-physical competitor of quantum communicators and its security is based on (out of Kirchhof's loop law) the Fluctuation-Dissipation Theorem, more generally, on the Second Law of Thermodynamics, indicating that the *security of the conceptual scheme is as strong as the impossibility of building a perpetual motion machine* (of the second kind).

First we briefly survey the foundations of the KLJN system [16,19]. Figure 3 shows the model of the idealized KLJN cypher designed for secure key exchange in [16]. The resistors $R_L$ and $R_H$ represent the low, $L$ (0), and high, $H$ (1), bits, respectively. At each clock period, Alice and Bob randomly choose one of the resistors and connect it to the wire line. The situation *LH* or *HL* represents secure bit exchange [16], because Eve cannot distinguish between them through measurements, while *LL* and *HH* are insecure. The Gaussian voltage noise generators (white noise with publicly agreed bandwidth) represent a corresponding thermal noise at publicly agreed effective temperatures $T_{eff}$ (typically $T_{eff} > 1$ billion Kelvins [18]). According to the *Fluctuation-Dissipation Theorem*, the power density spectra $S_{u,L}(f)$ and $S_{u,H}(f)$ of the voltages $U_{L,A}(t)$ and $U_{L,B}(t)$ supplied by the voltage generators in $R_L$ and $R_H$ are given by:

$$S_{u,L}(f) = 4kT_{eff}R_L \quad \text{and} \quad S_{u,H}(f) = 4kT_{eff}R_H \qquad (1)$$

respectively.

In the case of secure bit exchange (*LH* or *HL* situations), the power density spectrum of channel voltage $U_{ch}(t)$ and channel current $I_{ch}(t)$ are given as (see [16,19] for further details):

$$S_{u,ch}(f) = 4kT_{eff}\frac{R_L R_H}{R_L + R_H}, \qquad (2)$$

$$S_{i,ch}(t) = \frac{4kT_{eff}}{R_L + R_H}. \qquad (3)$$

Observe that during the *LH* or *HL* situation, due to linear superposition, Equation-2 is the sum of the spectra of two situations, i.e., when only the generator in $R_L$ is running:

$$S_{L,u,ch}(f) = 4kT_{eff}R_L\left(\frac{R_H}{R_L + R_H}\right)^2 \qquad (4)$$

and when the generator in $R_H$ is running:

$$S_{H,u,ch}(f) = 4kT_{eff}R_H\left(\frac{R_L}{R_L + R_H}\right)^2 \qquad (5)$$

The ultimate security of the system against passive attacks is provided by the fact that the power $P_{H \to L}$ with which resistor $R_H$ is heating $R_L$ is equal to the power $P_{L \to H}$ with which resistor $R_L$ is heating $R_H$ [16,19]. The proof can also be derived from Equation 3 for a frequency bandwidth of $\Delta f$ by:

$$P_{L \to H} = \frac{S_{L,u,ch}(f)\Delta f}{R_H} = 4kT_{eff}\frac{R_L R_H}{(R_L + R_H)^2} \qquad (6a)$$

$$P_{H \to L} = \frac{S_{H,u,ch}(f)\Delta f}{R_L} = 4kT_{eff}\frac{R_L R_H}{(R_L + R_H)^2} \qquad (6b)$$

The equality $P_{H \to L} = P_{L \to H}$ (see Equations 6) is in accordance with the *Second Law of Thermodynamics*; violating this equality would mean not only violating basic laws of physics and the ability to build a perpetual motion machine (of the second kind), but also that the eavesdropper (Eve) could utilize the voltage-current crosscorrelation $\langle U_{ch}(t)I_{ch}(t)\rangle$ to extract the bit [16]. However, $\langle U_{ch}(t)I_{ch}(t)\rangle = 0$ thus Eve have insufficient number of independent equations to determine the bit location during the *LH* or *HL* situation. The above *security proof against passive (listening) attacks, see situation c) in the Introduction*, holds only for Gaussian noises, which have the well-known property that their power density spectra or autocorrelation function provides the maximal information about the noise and no higher order distribution functions or other tools are able to serve with additional information.

Any deviations from this circuitry, including parasitic elements, inaccuracies, non-Gaussianity of the noise, etc. will cause potential information leak toward Eve.

To provide unconditional security against invasive attacks, including the man-in-the-middle attack, the fully armed KLJN cypher system, see Figure 4, is monitoring the instantaneous current and voltage values, at both ends (i.e., Alice and also Bob) [17-19], and these values are compared either via broadcasting them or via an authenticated public channel. The alarm goes off whenever the circuitry is changed or tampered with or energy is injected into the channel. It is important to note that these current and voltage data contain all the information Eve can have. This implies that Alice and Bob have full knowledge about the information Eve may have; a particularly important property of the KLJN system that can be utilized in secure key exchange.

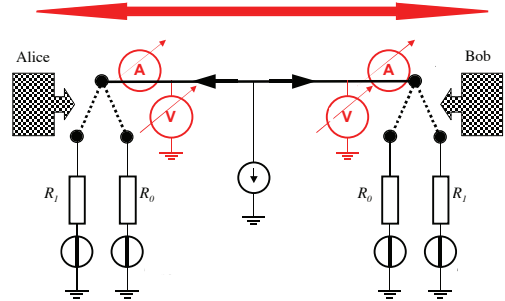

**Figure 4.** The KLJN wire communication arrangement. [3] To detect the invasive eavesdropper (represented for example by the current generator example at the middle), the instantaneous current and voltage data at measured at the two ends are broadcasted and compared. The eavesdropping is detected immediately, within a small fraction of the duration needed to transfer a single bit. Statistics of bit errors is not needed. The exchange of *even a single key bit is secure*.

This situation implies the following important features of the KLJN system [19]:

**1.** The KLJN system is *always secure*, even when it is built with non-ideal elements, in the following sense. The current and voltage data inform Alice and Bob about the exact information leak. Hence, they can always decide either to shot down the communication or to take the risk.

**2.** Even when the communication is jammed by invasive attacks or inherent non-idealities in the KLJN system, the system remains secure because no information can be eavesdropped by Eve without the full knowledge of Alice and Bob about this potential incidence, and without the full information Eve might have extracted, see the full analysis by Horvath in [19].

**3.** The KLJN system is naturally and fully protected against the man-in-the-middle attack [17], *even during the very first run of the operation* when no hidden signatures can be applied yet. This feature is provided by the unique property of the KLJN system that zero bit information can only be extracted during a man-in-the-middle attack because the alarm goes off before the exchange of a single key bit has taken place [17].

The outline of the prototype of the KLJN cypher [18] is shown in Figure 5. The various non-idealities have been addressed by different tools with the aim that the information leak toward Eve due to non-idealities should stay below 1% of the exchanged raw key bits. For the KLJN cypher it was 0.19% for the most efficient attack [18]. Here we briefly address two aspects of non-idealities:

**(i)** The role of the line filter (and that of the band limitation of the noise generator) is to provide the no-wave limit in the cable that is to preserve the core circuitry (see Figure 3) in the whole frequency band. That is, the shortest wavelength component in the driving noises should be much longer than the double of the cable length in order to guarantee that no active wave modes and related effects (e.g., reflection, invasive attacks at high frequencies, etc.) take place in the cable.

**(ii)** Another tool to fight non-idealities is the cable capacitance compensation (capacitor killer) arrangement (see Figure 5). With practical cable parameters and their limits, there is a more serious threat of the security: the cable capacitance shortcuts part of the noise current and that results in a greater current at the side of the lower resistance end yielding an information leak. This effect can be avoided by a cable-capacitor-killer [18] using the inner wire of a coax cable as KLJN line while the outer shield of the cable is driven by the same voltage as the inner wire. However, this is done via a follower voltage amplifier with zero output impedance. The outer shield will then provide all the capacitive currents toward the ground and the inner wire will experience zero parasitic capacitance. Without capacitor killer arrangement and practical bare-wire line parameters, the recommended upper limit of cable length is much shorter depending on the driving resistor values $R_L$ and $R_H$.

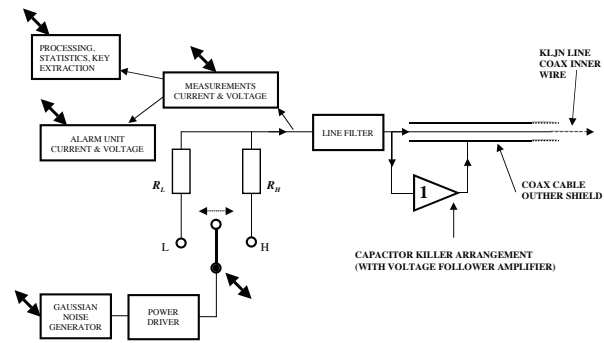

**Figure 5.** A practical KLJN cypher [18]. Double-ended arrows symbolize computer control.

## 3. Security proofs and attacks

The idealistic system is absolutely secure but real systems are rarely idealistic. Several hacking attack types were published based on the non-ideality of circuit elements causing deviations from the idealistic circuitry, see discussions in [20-25]. Each of these attacks triggered a relevant security proof that showed the efficiency of the defense mechanism (see Figure 4). Also, all attack types were experimentally tested by Mingesz et al [18] and the theoretical security proofs were experimentally confirmed. At practical conditions, the most effective attack utilized non-zero wire resistance, even though in that paper [20] by Scheuer and Yariv, calculation errors were made and the real effect was 100 times weaker, as shown later by Kish and Scheuer [21]. Other attack types of practically less significance were by: Hao [23], based on differences in noise temperatures (theoretical response: [24], experimental one: [18]); Mingesz *et al.* [18] and Kish [24], using deviations of resistance values of the resistor pairs at the two ends, Kish *et al.* [19], using wire capacitance and inductance (response in [19]); and Liu [25], exploiting delay effects (response: [19]). The level of allowed information leak is the choice of Alice and Bob and its actual value is given by the invested resources to reach ideality and it depends on how much speed is given up for that purpose. In the experimental demo [18] the strongest leak was due to wire resistance, where 0.19% of the bits leaked out to Eve and the fidelity of key exchange was 99.98%. This is a very good raw bit leak, which can be completely removed by a simple privacy amplification. [26].

**4. Privacy amplification**

Privacy amplification is a technique to ensure the security of an encryption scheme of which the key has been partially exposed. Horvath, et al [26] studied a simple privacy amplification: XOR-ing subsequent pairs of the key bits, thereby halving the key length, while exponentially reducing the information leak. They found that in contrast to quantum key distribution schemes, the high fidelity of the raw key generated in the KLJN system allows the users to *always sift a secure shorter key* if they have an upper bound on the eavesdropper probability to correctly guess the exchanged key bits. The number of privacy amplification iterations needed to achieve information leak of less than $10^{-8}$ in the case of the 0.19% information leak is two, resulting in a corresponding slowdown by a factor of four.

**5. Securing computers and hardware**

An important advantage [27] of the KLJN method, that is impossible to achieve with quantum communicators is that the KLJN circuitry can be integrated on computer chips to provide secure key exchange for secure data communication between hardware units within a computer. Due to the short wires in a computer, the main non-idealities [19] (wire resistance, inductance and capacitance) are negligible, thus the key exchange can run under idealistic conditions to provide unconditional security without further processing [27], such as privacy amplification.

**6. High-speed secure key distribution over chain networks**

In [28], instead of point-to-point secure key exchange, a high efficiency, secure chain network was proposed. Each agent had two KLJN systems and one was used to do exchange in the left direction in the chain and the other one in the right direction. The whole network consists of two parallel networks: i) the chain-like network for secure key exchange with left and right neighbors; and ii) a regular non-secure internet network with a Coordinator-server. The protocol provides a teleportation-type multiple telecloning of the key bit because the information transfer can take place without the actual presence/recognition of the information bit at the intermediate points of the chain network. At the point where telecloning takes place, the clone is created by the product of a bit coming from the regular network and a secure bit from the local KLJN ciphers. For details see [28].

**7. Utilizing wire lines in use: power lines, phone lines, etc [29]**

Since existing wires, such as power lines connect all buildings, apartments and offices, they have great potential to be utilized for secure key exchange through the KLJN method. However, there is a problem: as we have seen in Section 2, to maintain security and to avoid that the voltage/current alarm goes off, the core circuitry of the KJLN secure key exchange system must be preserved. When we plan to utilize existing lines for realizing secure key distribution over networks, filters can be used to create distinct isolated regions of the wire that can be used for the core circuitry. Without the limitation of generality, the examples below will utilize power lines for this purpose.

7.1 The filter method for single lines

Figure 6 [29] shows an example that uses a KLJN frequency Band Excluder (BE) and Band Pass (BP) filters to preserve a single Kirchhoff loop's characteristics in the KLJN frequency band between two KLJN communicators with one intersection between them. Both, the original non-KLJN load (power consumer, phone, or internet card represented by $R_N$) and the KLJN communicators will work using their own, non-overlapping frequency bands. BE excludes the KLJN frequency band and passes all other important frequencies. BP passes the KLJN frequency band and excludes all other frequencies. BP filters must be used in any case at the communicator outputs to avoid transient and eavesdropping probing signals out of the KLJN band.

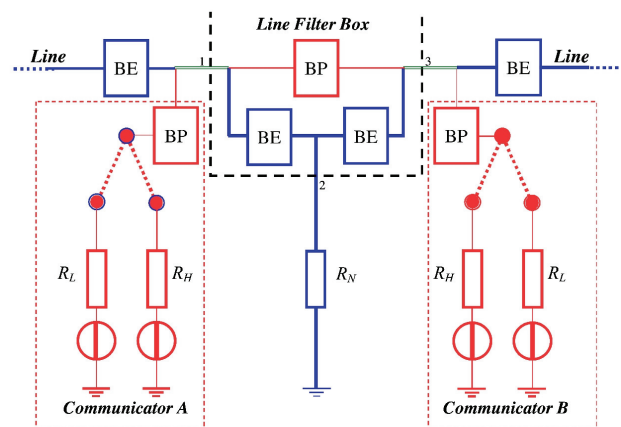

**Figure 6.** Example for how to use KLJN frequency Band Excluder (BE) and Band Pass (BP) filters to preserve a single Kirchhoff loop in the KLJN frequency band between two KLJN communicators with one intersection between them [29]. Thick (blue, non-numbered) lines: original line current; thin (red, non-numbered) lines: KLJN current; thick (green) lines 1 and 3: both types of currents.

Figure 7 shows that, though the topology in Figure 6 may look complex, the circuitry in the Line Filter Box in Figure 6 can be contained by a box with only three electrodes. If there is more than one consumer along the line between the communicators, each of them must be fed via a separate Line Filter Box.

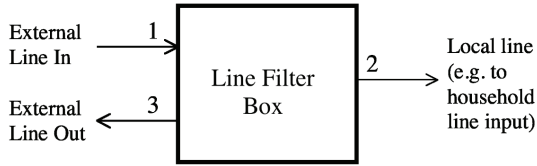

**Figure 7.** The line filter box (see Figure 1) should be installed at each intersection of the line to separate the non-KLJN communicator loads from the KLJN frequency band [29].

7.2 Ground-level method with 3-phase power lines

In the case of 3-phase power lines [29], we can make use of the fact that, in the idealized case, when the load is symmetric on the phases, there is no current flowing from the common point of the 3-phase transformers to the ground. Then the KLJN communicators can be connected between the common points of the 3-phase transformers and the ground, see Figure 8. This arrangement can help to reduce the problem of working with high voltages because the common point of the transformers is at ground potential in the idealized case.

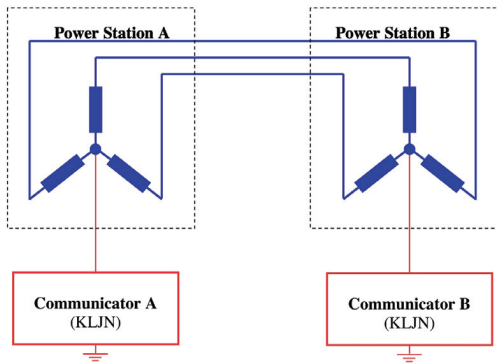

**Figure 8.** Communication via idealized 3-phase power lines with symmetric loads of the 3-phase transformers at Power Stations A and B, respectively [29].

However, in practice, there are non-idealities and asymmetries among the lines (load, phase, etc.) therefore filters will be necessary to avoid problems. Figure 9 shows a possible solution of such problems. In the KLJN frequency range the filters drive the current through the communicators and out of that frequency range the current goes into the ground. Note, if there is an intersection of cables with asymmetric loads between Power Stations A and B then filters must be used similarly to Figure 6 and then the problems with high voltages cannot be avoided. Therefore, the arrangements in Figures 8 and 9 are most advantageous when there is no intersection between the two power stations.

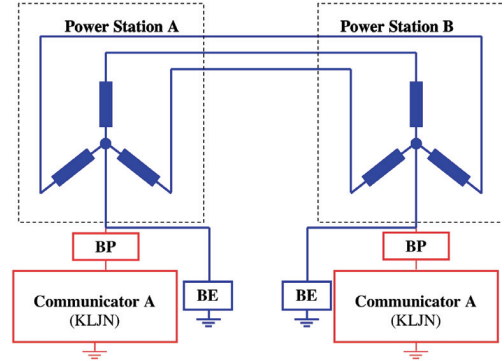

**Figure 9.** A more robust version [29] of the scheme shown in Figure 8.

**Summary**

After introducing the basic concept and characteristics of the KLJN key exchange, we focused on the practically important aspect of utilizing existing wire lines, such as power lines. We have shown with some simple circuit demonstrations that the KLJN units can use existing wires for communication. The filter method can also be used with phone and internet lines and for getting the KLJN current around switchers and multiplexer units. The concrete realization and further development of these ideas is straightforward, though not trivial, but we will not go in details since it is out of the scope of the present paper.

Finally, we must answer the following questions. Is the communication still secure if the eavesdropper removes a filter or if a non-filtered new intersection is made on the line? The answer is straightforward: the communication becomes non-secure; however the current-voltage alarm system (see Figure 4), which is comparing the KLJN voltages and currents at the two units, will go off and the communication will immediately be terminated for security reasons. Therefore the eavesdropper cannot extract any useful bits of information.

**Acknowledgments**

The Authors are grateful to Dr. Makoto Naruse for discussions and his inputs. LBK is grateful for related earlier discussions with Zoltan Gingl, Robert Mingesz and Tamas Horvath. This paper was supported in part by Grant-in-Aid for Scientific Research funded by Japan Society for the Promotion of Science.

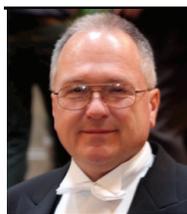

**Laszlo B. Kish** (1955-1999: Kiss), physicist, has been doing research on a wide scale of phenomena and applications concerning stochastics; noise and fluctuations,; physical informatics; sensing; and materials science. He is full professor of electrical and computer engineering at Texas A&M University.

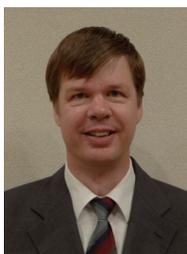

**Ferdinand Peper** received both his M.Sc. (1985) and Ph.D. (1989) from Delft University of Technology, the Netherlands, in Theoretical Computer Science. He is a Senior Researcher at the National Institute of Information and Communications Technology (NICT), and he is a member of IEEE and ACM. His research interests include emerging research architectures, cellular automata, nanoelectronics, fluctuation-driven computation, neuron science, and networking.